\documentclass[aps,prl,reprint,twocolumn,showpacs,letterpaper]{revtex4-1}
\usepackage{graphicx}
\usepackage{dcolumn}
\usepackage{bm}
\usepackage{xcolor}
\usepackage{multirow}
\usepackage{subfigure}
\usepackage{hyperref}
\usepackage{amssymb}
\usepackage{color}
\usepackage{graphics}\usepackage{epsfig}\usepackage{subfigure}
\usepackage{footnote}
\usepackage{amsmath}

\begin{document}
\title{Phonon-assisted  magnetic Mott-insulating state\\ in the
charge density wave phase of single-layer 1TNbSe$_2$}
\author{Matteo Calandra}
\affiliation{Sorbonne Universit\'e, CNRS, Institut des
  Nanosciences de Paris, UMR7588, F-75252, Paris, France}

\begin{abstract}
We study the structural, electronic and vibrational properties of 
single-layer 1TNbSe$_2$ from first principles. 
Within the generalized gradient approximation, the 1T polytype is
 highly unstable with respect to the 2H. The  DFT+U method improves
 the stability of the 1T phase, explaining its detection in experiments.
 A charge density wave occurs with a
$\sqrt{13}\times\sqrt{13}~R30^{\circ}$ periodicity, in
agreement with STM data. 
At $U=0$, the David-star reconstruction displays a flat band below the
Fermi level
with a marked d$_{z^2-r^2}$ orbital character of the central Nb.
The Hubbard interaction induces a  magnetic 
Mott insulating state. Magnetism distorts the lattice around the central
Nb atom in the star, reduces the hybridization
between the central Nb d$_{z^2-r^2}$ orbital and the  
neighbouring Se p-states and lifts in energy the empty d$_{z^2-r^2}$ flat band 
becoming non-bonding. 
This cooperative Jahn-Teller and correlation
effect amplifies the Mott gap. 
Our results are relevant for the broad class of
correlated insulator in the presence of a strong Jahn-Teller effect.
\end{abstract}

\maketitle

The 2H polytypes of transition metal dichalcogenides (TMDs) like
2HNbSe$_2$, 2HTaS$_2$ or 2HTaSe$_2$, are ideal materials to study
the interplay of superconductivity and charge-density wave (CDW) \cite{Disalvo_rev}, 
as the electron-electron interaction has a minor 
role and metallicity survives in the CDW state. 
1T polytypes, such as 1TTiSe$_2$,
1TTaS$_2$ or 1TTaSe$_2$, are substantially different and more critical, 
as in the CDW ground state the electron-electron interaction is
supposed to substantially affect the electronic structure
\cite{Rossnagel,Bianco,Darancet} and lead to
Mott\cite{Fazekas}, Slater or magnetic insulating states and even to spin
liquids \cite{Sliquid}. While 1T-TaSe$_2$ is metallic at low temperature, the CDW
state of 1T-TiSe$_2$ seems to have a very small electronic gap ($~0.1$
eV).  The nature of the low
temperature state of 1T-TaS$_2$ is still
very controversial as it has been proposed to be a Mott
insulator\cite{Fazekas}, an Anderson disordered
insulator\cite{Darancet}
or, more recently, a metal\cite{Hoffman_kz}.

\begin{figure*}
\begin{minipage}[c]{0.3\linewidth}
\includegraphics[width=5cm]{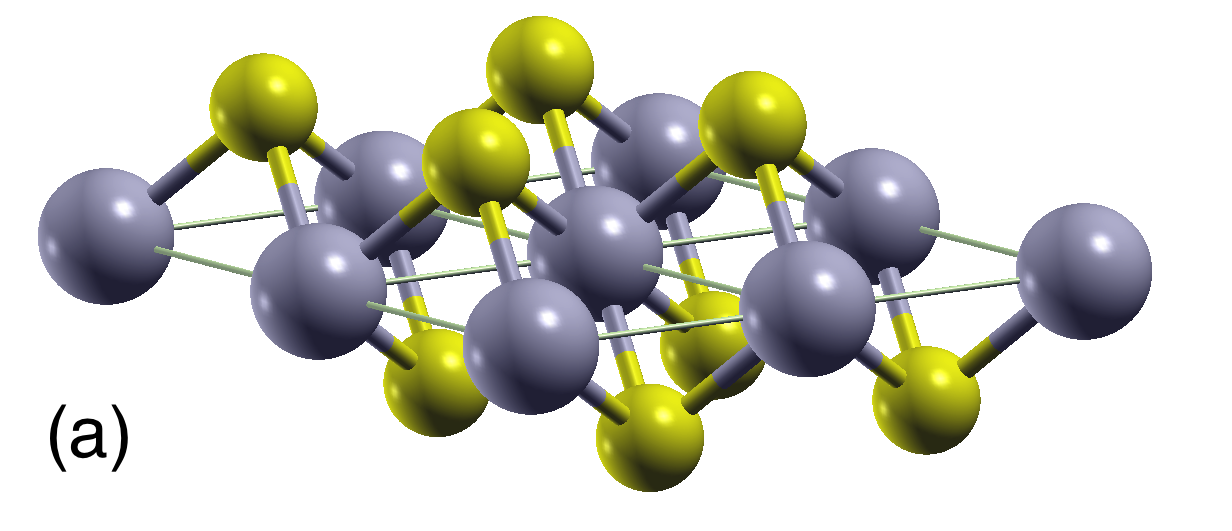}\\
\includegraphics[width=5cm]{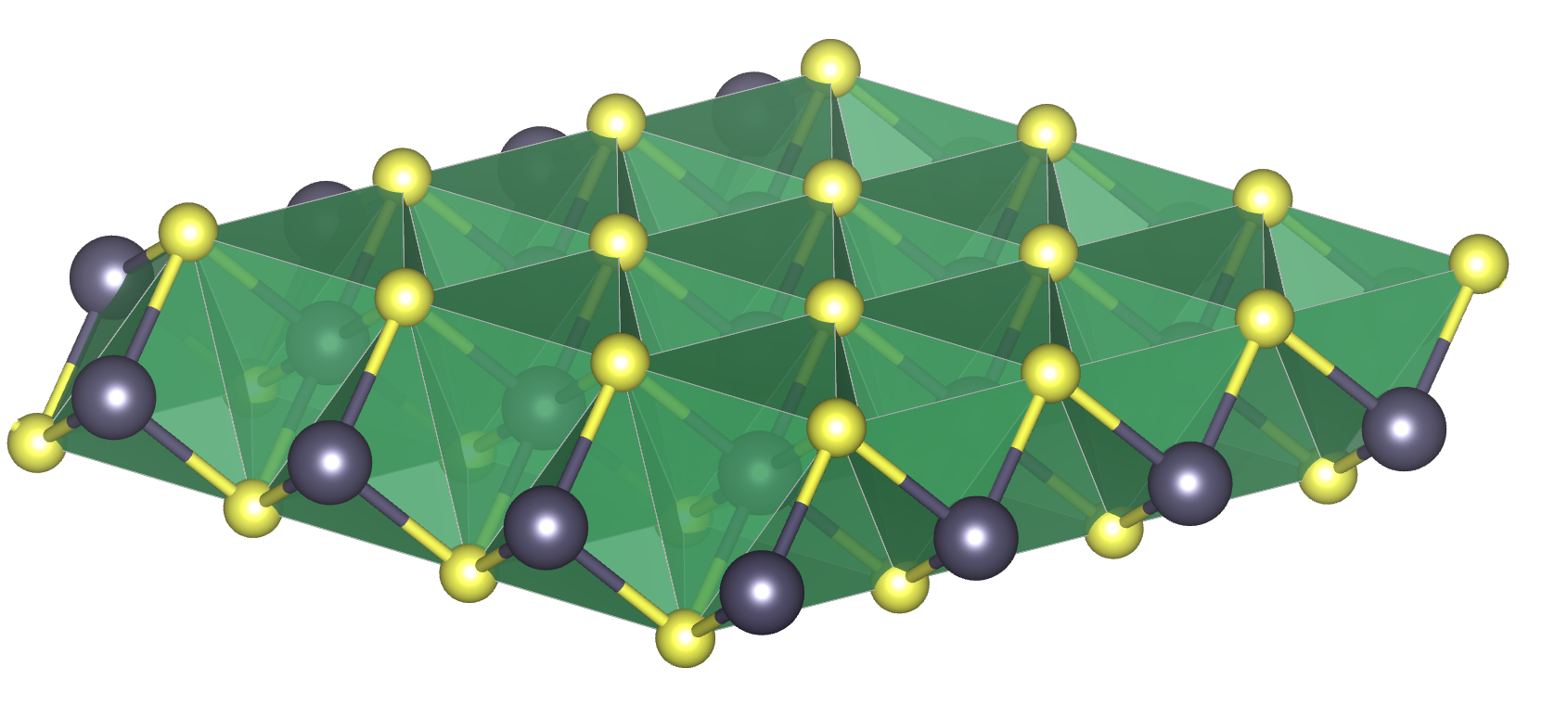}
\end{minipage}\hfill 
\begin{minipage}[c]{0.7\linewidth}
\includegraphics[width=5.6cm]{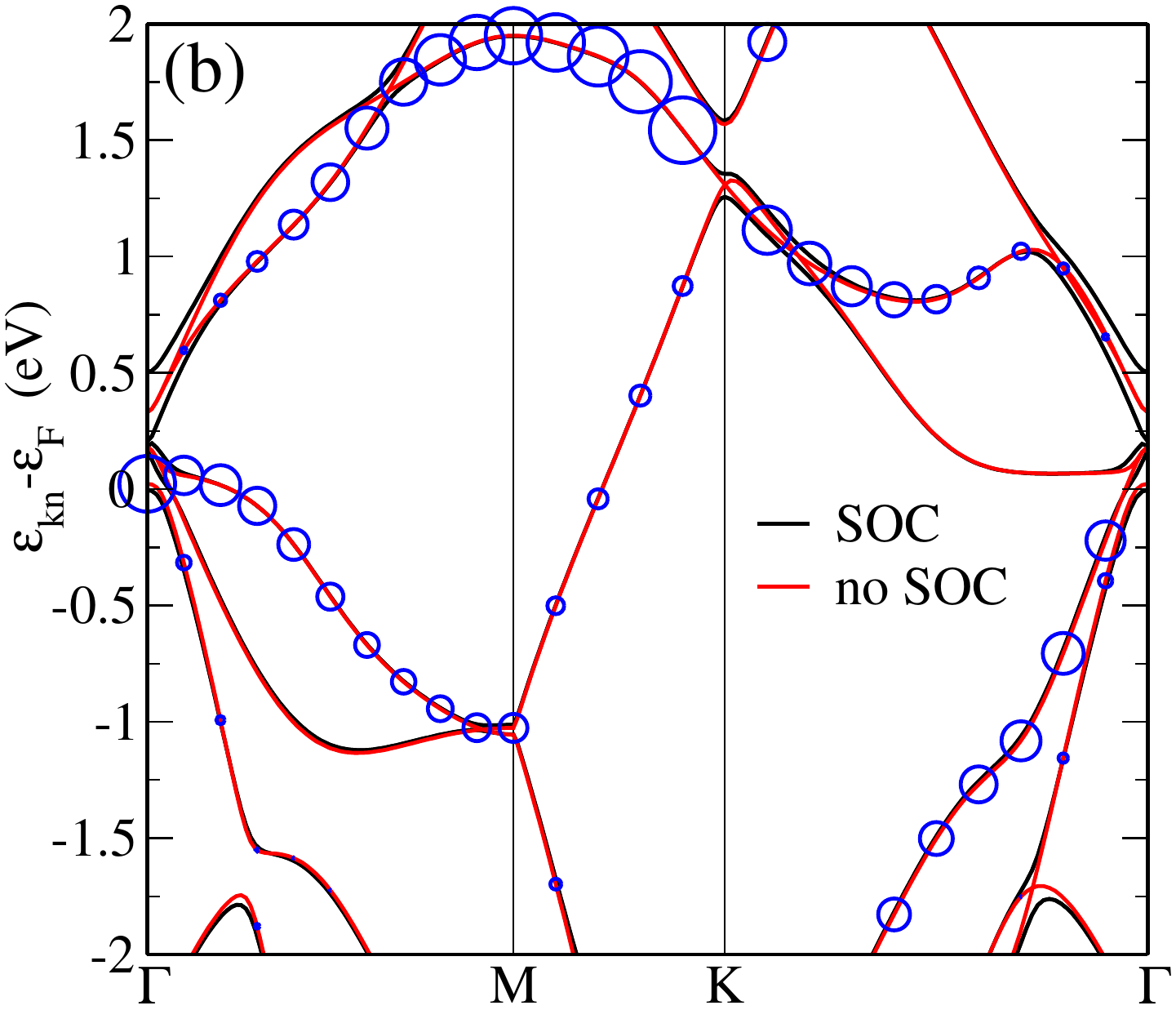}
\includegraphics[width=5.5cm]{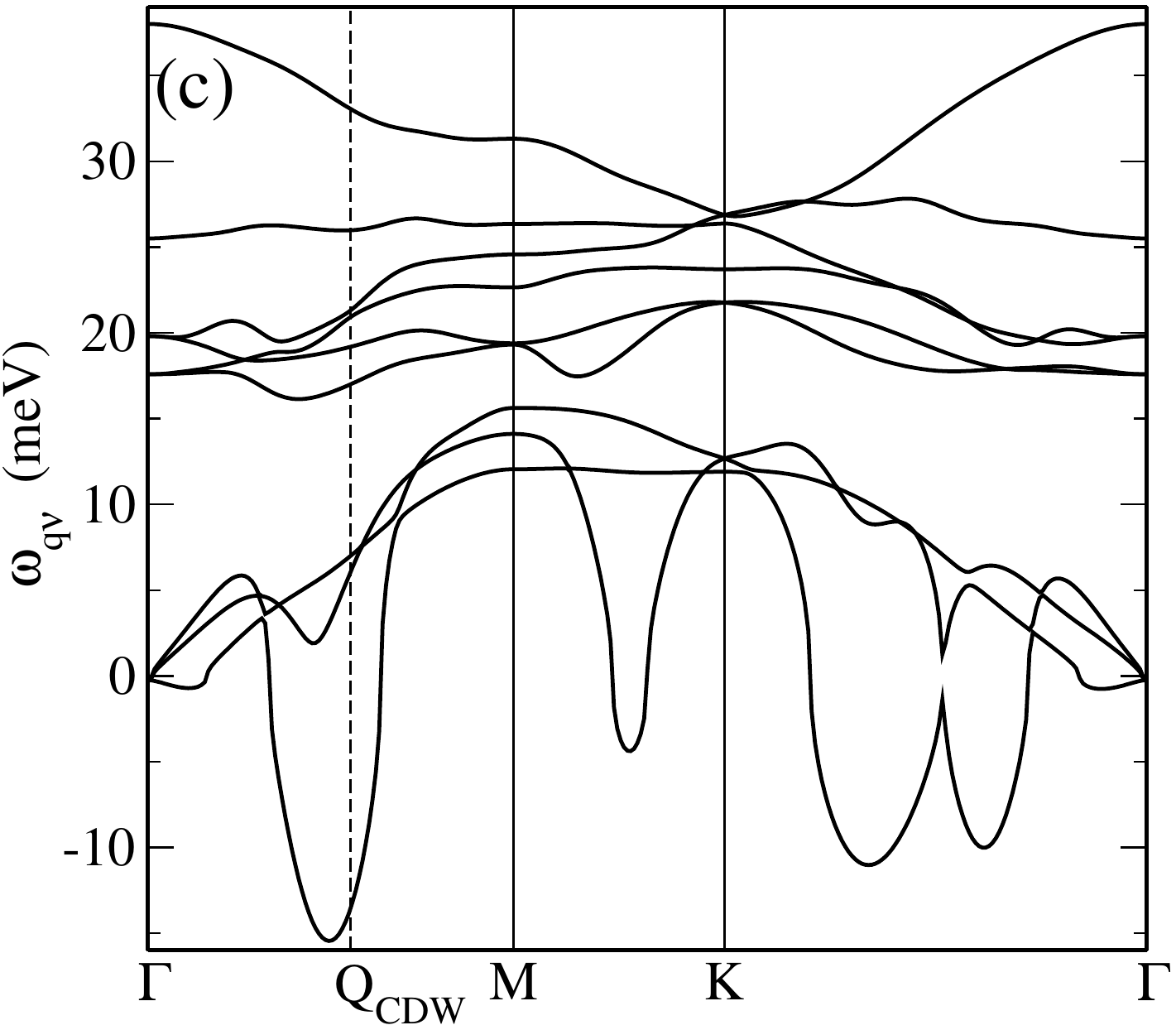}
\end{minipage}
\caption{(a) Crystal structure of an undistorted 1TNbSe$_2$
  single-layer (Nb atoms in grey, Se atoms in yellow). In the bottom picture the octahedral coordination 
  around the Nb atom is shown. Electronic structure (b) and phonon 
  dispersion (c) of single-layer 1TNbSe$_2$  in the high-symmetry phase. 
In (b) SOC (noSOC) means with (without)   inclusion of spin-orbit coupling.
In the non-relativistic case, the size of the blue dots is proportional to the 
3d$_{z^2-r^2}$ orbital component (at ${\bf \Gamma}$
  the larger dot correspond to $53\%$ d$_{z^2-r^2}$ component).  }\label{Fig:undist}
\end{figure*}

The electronic structure of high-T phase of 1T polytypes in the
absence of a CDW, (hereby
labeled high-symmetry structure) can be understood, to some extent, by
neglecting interlayer coupling and invoking the crystal-field symmetry 
around the transition metal (TM)\cite{Canadell,Mattheiss}.
The chalcogene atoms form a slightly distorted octahedron
around the TM, as shown in Fig.\ref{Fig:undist} (a). In the case of an
 undistorted 
octahedron, the
atomic d-levels are splitted into triply degenerate t$_{2g}$ orbitals (d$_{x^2-y^2}$,
d$_{z^2-r^2}$,$d_{xy}$) and doubly degenerate e$_g$ orbitals (d$_{xy}$,
d$_{xz}$) at higher energy. The distortion of the octahedron breaks the 
degeneracy of the t$_{2g}$ manyfold and 
lowers the energy of the  d$_{z^2-r^2}$ orbital. For the
case of Ta or Nb, the nominal
d$^1$ valence  leads to an half-filled
d$_{z^2-r^2}$ state at zone center (see Fig. \ref{Fig:undist}(b)), with the exact position of the Fermi level
depending on the hybridization between the chalcogene and the TM
orbitals. 

In the case of a $\sqrt{13}\times\sqrt{13}~R30^{\circ}$
CDW with the formation of a so-called
David star clustering, each one of the 13  d$_{z^2-r^2}$ states involved in the star
carries one electron, leading
to a formally half-filled HOMO molecular state. The narrowness
of the HOMO state, related to the weaker inter-star interaction, is
prone to electronic instabilities. Indeed it has been
suggested\cite{Fazekas} that bulk 1TTaS$_2$ is a Mott insulator.
The experimental validation of this scenario in 1TTaS$_2$ has proven
to be very controversial mainly due to the experimental incertitude in the determination 
of the interlayer stacking\cite{Wiegers,Tanda,Nakanishi,Ritschel} and
the consequent orbital ordering affecting hopping along the $c-$axis.
Moreover, the occurence of Jahn-Teller distortion
within $\sqrt{13}\times\sqrt{13}~R30^{\circ}$ pattern emerging from
the stabilization of a spin $1/2$ magnetic
state on each star has never been discussed. 
Finally, recent ARPES experiments\cite{Hoffman_kz}
display a metallic band dispersion along k$_z$,
meaning that the ground state of 1TTas$_2$ is still very
controversial.

The validity of the Mott scenario could be addressed by synthesizing
1TTaS$_2$ single-layers, however this attempt has been, for the moment, unsuccesful as
(i) single-layer TaS$_2$ grown epitaxially on Au(111)  assumes
the 2H polytype\cite{Sanders} and (ii) preliminary results on exfoliated suspended
samples seems to show that 1TTaS$_2$ single-layer undergoes a different kind
of CDW order\cite{Sakabe}. Although some supported samples 
shows Raman spectra consistent with a
$\sqrt{13}\times\sqrt{13}~R30^{\circ}$ CDW
\cite{Albertini}, nothing is known on the nature of their
electronic structure. 

A way out of this impasse is to consider 1TNbSe$_2$ single-layer. 
Although in bulk form and in most exfoliated samples NbSe$_2$ 
assumes the 2H polytype, it has been recently shown that  
single-layer of 1T polytype can be grown epitaxially on bilayer
graphene\cite{Nakata}, keeping the substate at temperatures larger than
 500 K. Scanning tunneling spectroscopy shows the occurrence
of a $\sqrt{13}\times\sqrt{13}~R30^{\circ}$ CDW. ARPES and STS 
measurements in the CDW phase
seems to show the occurrence of
an $\approx 0.3-0.4$ eV gap at zone center. No measurements of the band structure
in other regions of the Brillouin zone are at the moment available.
Not surprisingly, the ARPES data in the CDW phase are in stark disagreement with 
electronic structure calculations in the high symmetry 1TNbSe$_2$ phase\cite{Nakata}.
Finally, and most
interesting, Bischoff {\it et al.}\cite{Bischoff} recently showed that it is possible
to transform the surface of exfoliated 2HNbSe$_2$ samples 
into a 1TNbSe$_2$ phase by 
applying STM bias pulse of $4$V for $100$ ms. The 1TNbSe$_2$
phase in the top layer is then (meta)stable at $77$ K and 
displays the occurrence of the
characteristic   $\sqrt{13}\times\sqrt{13}~R30^{\circ}$
reconstruction. 

In this work, by using electronic structure calculations\cite{QE1,QE2}, 
we show that single-layer 1TNbSe$_2$ 
undergoes a CDW instability with a
$\sqrt{13}\times\sqrt{13}~R30^{\circ}$ reconstruction, in agreement
with experiments. 
A cooperative correlation and Jahn-Teller effect
stabilizes a spin-$1/2$ magnetic Mott-insulating state in
reduced dimension.

We first consider the high-symmetry 1TNbSe$_2$ structure shown in
Fig.\ref{Fig:undist}. We use the in-plane experimental lattice parameter
( $a_{\rm Exp.}=3.44 \AA$) and  minimize the internal coordinates. 
More technical details are given in
the supplemental materials\cite{Supplemental}.
The calculated electronic structure shown in
Fig. \ref{Fig:undist} (b)  is similar to that of the undistorted 1TTaS$_2$
monolayer\cite{Darancet}, however the hybridization between chalcogene and TM
d-orbitals is much stronger in this case so that the bottom of the
$d_{z^2-r^2}$ band is substantially lower in energy and there is no
gap between Nb d and Se p states. Finally, as the 1T structure breaks inversion
symmetry, we carry out a fully relativistic electronic structure
calculation,  but we found the spin-orbit coupling  to be negligible for the
undistorted phase, as shown in Fig. \ref{Fig:undist} (b). For this reason we
neglect relativistic effects in the rest of the paper.

We calculate the vibrational properties of the
high symmetry 1TNbSe$_2$ phase and, not surprisingly, we find strongly unstable phonon
modes (Fig. \ref{Fig:undist} (c) ) with the largest instability very close to the ordering
vector, {\bf Q}$_{CDW}=2\pi/\sqrt{13}a(1,0)$, that is the nearest vector along
$\Gamma M$ compatible with a $\sqrt{13}\times\sqrt{13}$ distortion
(see Ref. \cite{Amy}, Fig. 1 right).
\begin{figure*}[t]
\includegraphics[width=0.4\linewidth]{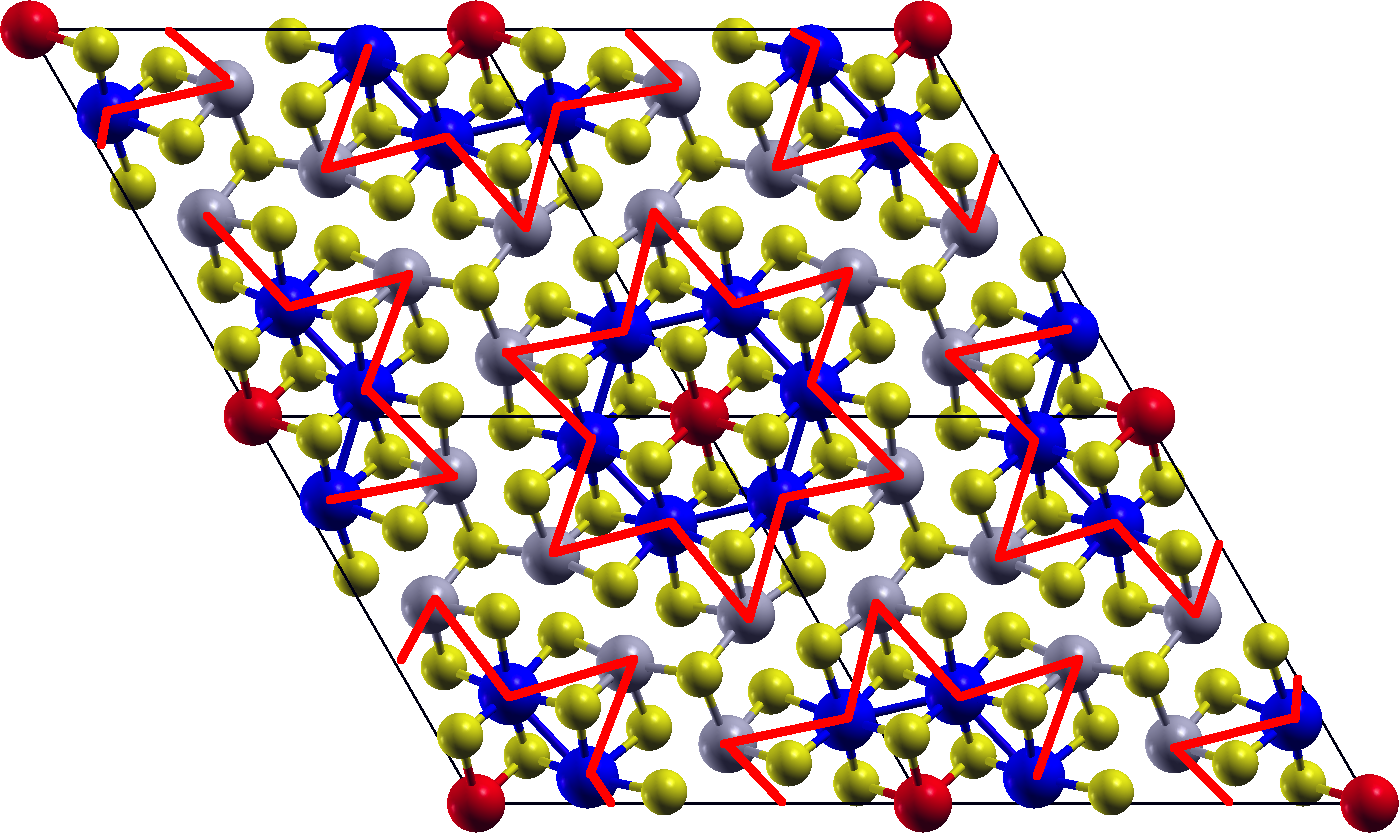}
\includegraphics[width=0.3\linewidth]{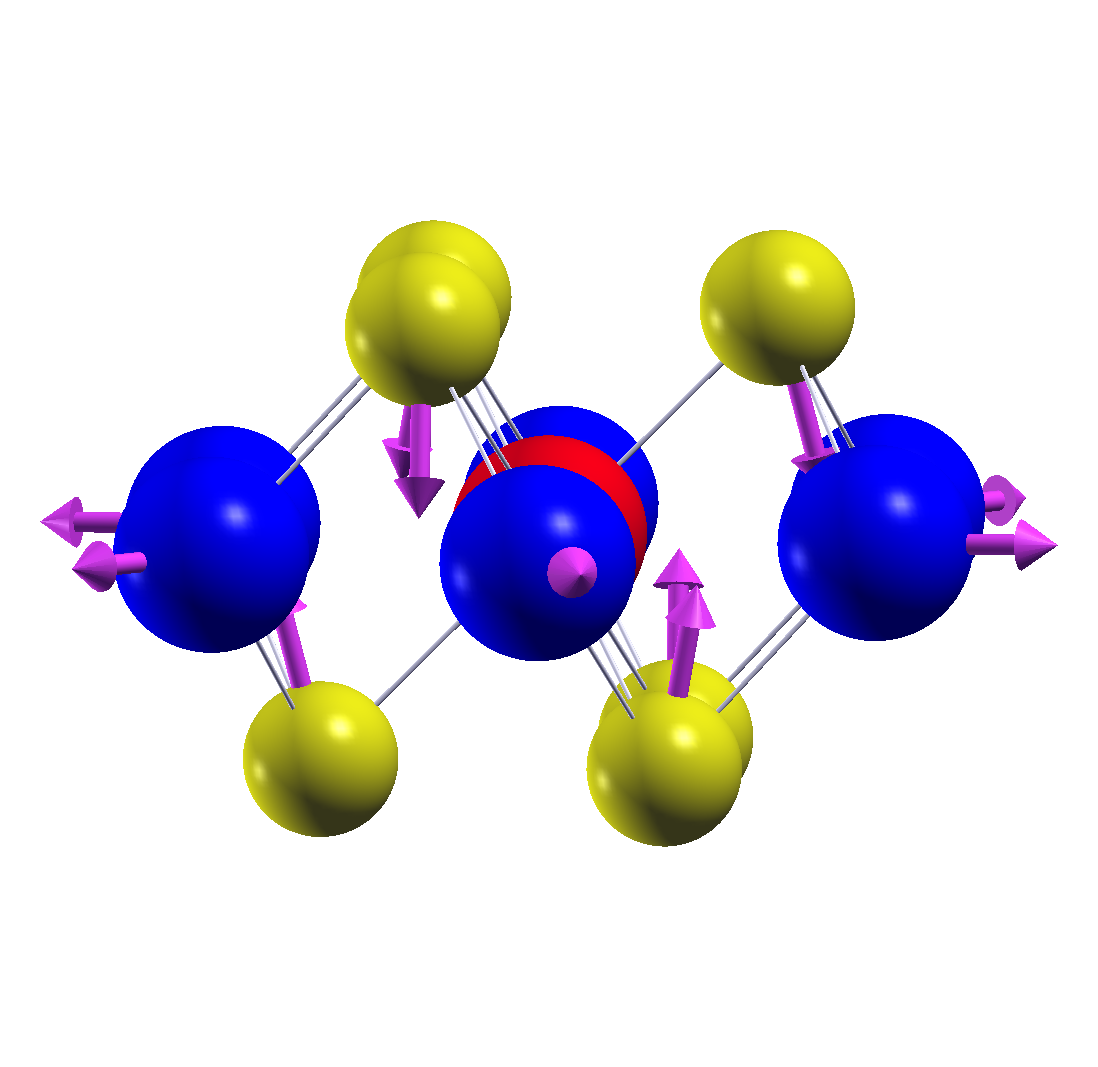}
\includegraphics[width=0.27\linewidth]{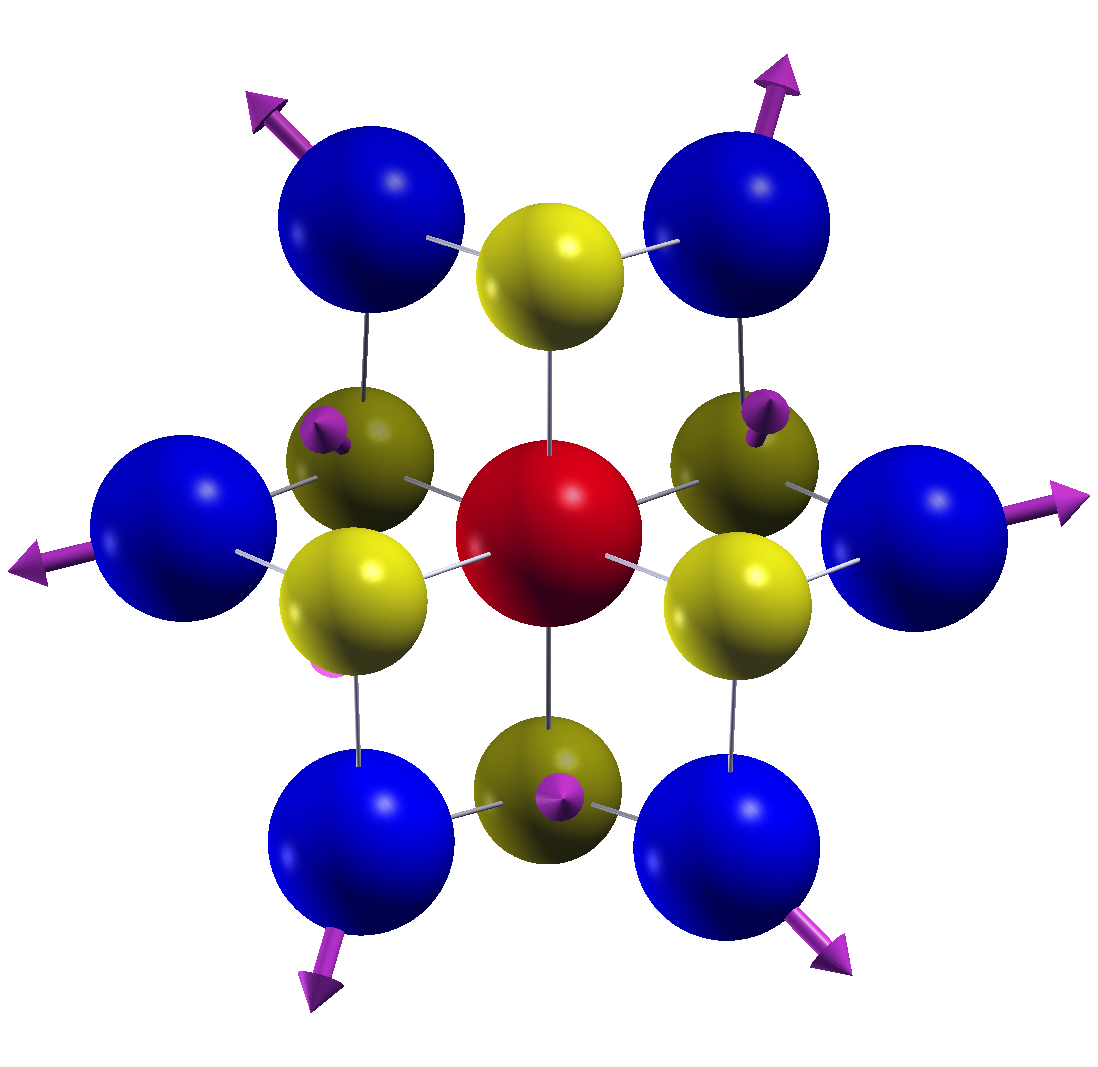}
\caption{Left: Optimized crystal structure for the low temperature phase of 
  1TNbSe$_2$. There are three inequivalent Nb sites: central Nb (red),
hexagonal site belonging to the $\sqrt{7}\times\sqrt{7}$ cluster 
(blue) and peripheral Nb sites (grey). The Se atoms are shown in 
yellow. The red line is a guide to the eye to recognize the 
$\sqrt{13}\times\sqrt{13}$ cluster (David star). The inclusion of the Hubbard term leads to a ferromagnetic state with 
an important local distortion around the central Nb atom in the star. However the difference 
between the ferromagnetic and non-magnetic structures are not visible on 
this scale.  
The magnetic state has a 
total magnetization of $+1\mu_B$. The magnetic 
moments on Nb atoms are:
$0.58\mu_B$ central (red) Nb, $0.0417$ blue Nb, $-0.0256$ peripheral 
(grey) Nb. The moduli of the magnetic moments on Se atoms are smaller 
than $0.08 \mu_b$. Center and left: magnetic induced distortion (side 
and top views) in the 
central $\sqrt{7}\times\sqrt{7}$ cluster. The central atom is 
unshifted. 
The 6 blue Nb atoms interacting ferromagnetically with the central 
one, moves apart in-plane. The out-of-plane Se approaches the Nb
basal plane.
In the top view (rightmost panel) the bottom Se atoms are shown in a
darker yellow color. }\label{Fig:struct_dist}
\end{figure*}     
We then consider a $\sqrt{13}\times\sqrt{13}R30^{o}$ supercell
and perform structural optimization assuming a non-magnetic state. 
We obtain the David-star structure depicted
in Fig. \ref{Fig:struct_dist} and reported in the supplemental
materials\cite{Supplemental}. The structure has three inequivalent Nb sites, central
(red), hexagonal site belonging to the $\sqrt{7}\times\sqrt{7}$
cluster (blue) and peripheral (grey). 
The energy gain of the CDW structure with respect to the
 non-magnetic high-symmetry 1T structure is very large (4.2 mRyd/Nb), however
the CDW structure is still substantially disfavoured from the 2H
high-symmetry structure (see Tab. \ref{tab:Energy}).
Thus, on the basis of non-magnetic calculations, it would be very
difficult to detect any 1TNbSe$_2$ polytype. 

\begin{figure*}
\includegraphics[width=0.32\linewidth]{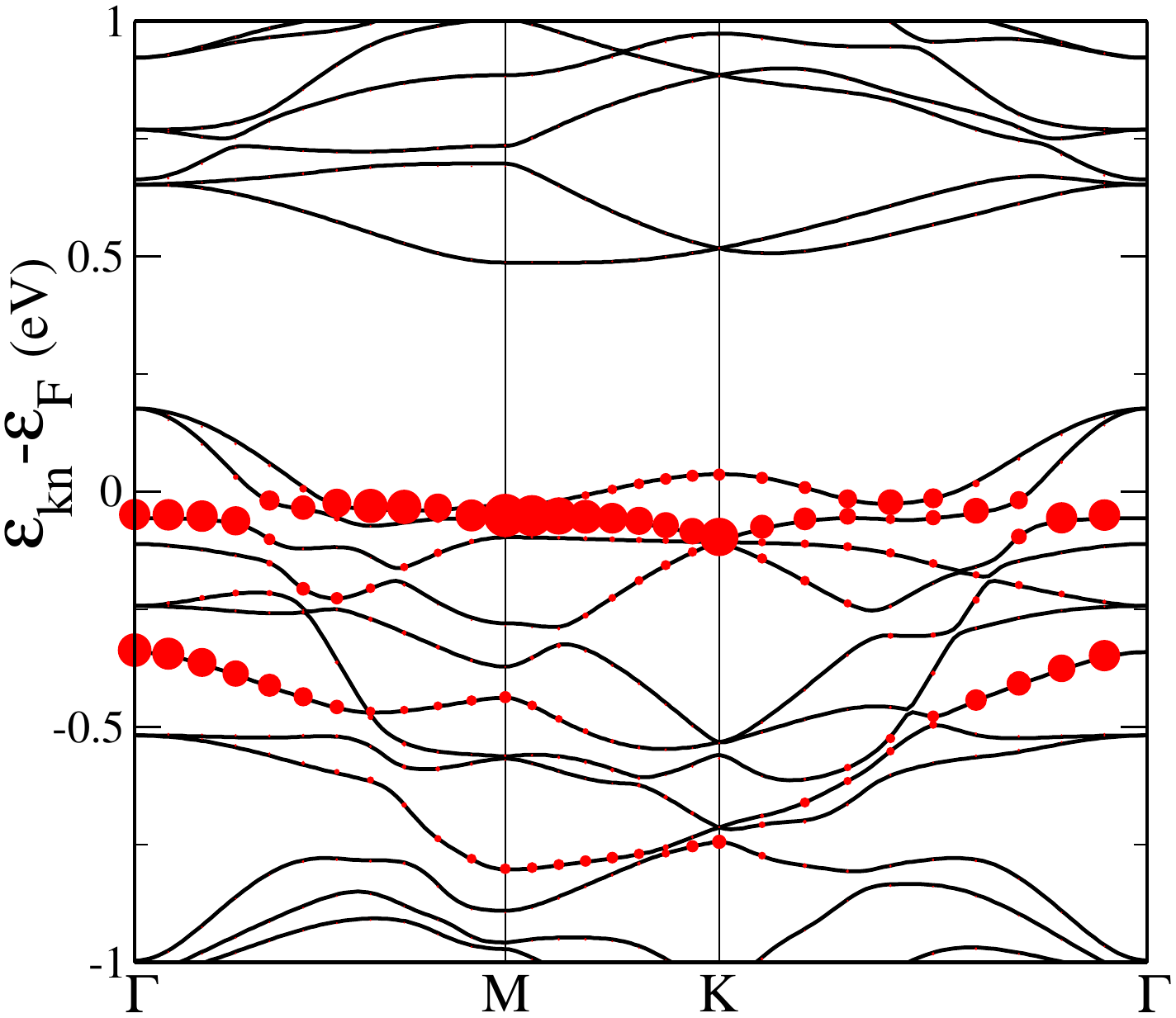}
\includegraphics[width=0.32\linewidth]{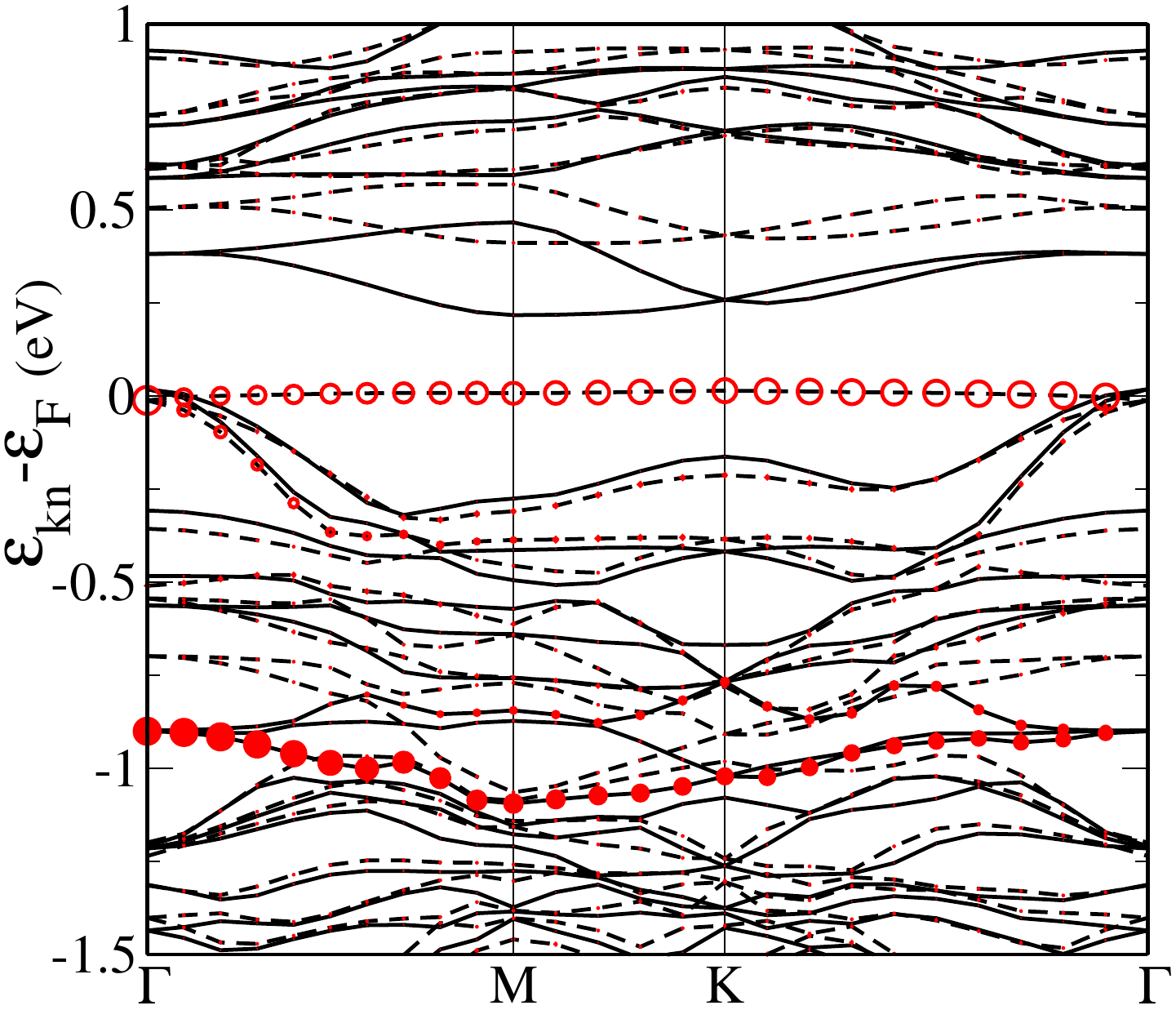}
\includegraphics[width=0.32\linewidth]{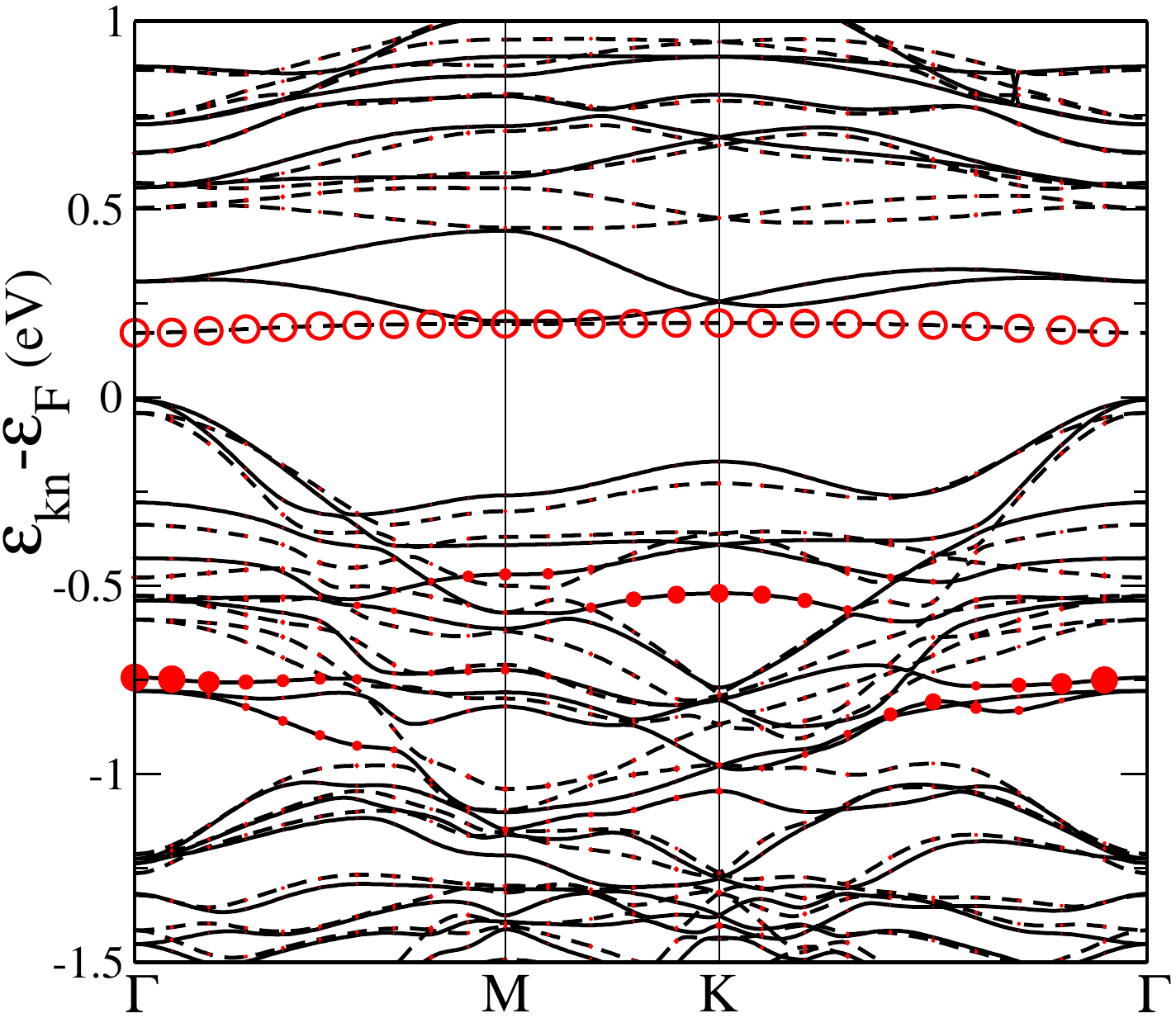}
\caption{Left: Electronic structure in the charge density wave phase 
  of 1TNbSe$_2$ within the non-magnetic GGA. The size of the red dots is proportional to the 
  Nb central-atom $d_{z^2-r^2}$ orbital character, (the red dot at zone 
  center and $-0.05$ eV corresponds to $15\%$ orbital 
  character). Center: same as left but in the GGA+U approximation on 
  top of the non-magnetic geometry. Continous (dashed) lines 
  are majority (minority)  spin bands. The size of the full (open) 
  dots is proportional 
to the Nb central-atom $d_{z^2-r^2}$ up-spin (down-spin) orbital 
character. Right: as in the center panel, but on top of the GGA+U 
geometry. 
The Hubbard induced distortion raises in energy the minority spin band and 
increases Mott gap.}\label{Fig:bands}
\end{figure*}

Despite the large number of atoms in the cell, the electronic
structure at the Fermi level is fairly simple. The top of the
valence band is at zone center and is composed mostly of Se states
originating from all the chalcogenes in the star. At the Fermi level, 
there is a very flat band having a marked 
 d$_{z^2-r^2}$ character from the central Nb atom (red dots in
 Fig. \ref{Fig:bands} ). 
The band has several avoided crossings due to
 the reduced symmetry of the CDW state. As it will be soon clear, tracking the 
central-atom d$_{z^2-r^2}$ orbital-component allows to detect
the upper and lower Hubbard bands. 

The narrowness of the band suggests the possible occurrence of
electronic instabilities. Then, in an effort to stabilize a magnetic
state, we peform a  spin-polarized
generalized gradient approximation on top the GGA\cite{PBE} non-magnetic
geometry. However, we always converge to a non-magnetic state.
We then consider the GGA+U approximation as implented in Ref. \cite{Cococcioni}.
We calculate the Hubbard U self-consistently \cite{CococcioniSCF}
and obtain $U=2.95$ eV, slightly larger of
what has been obtained in 1TTaS$_2$ \cite{Darancet}. 
We then perform magnetic calculations within this approximation.
We now find that the ground state of single-layer 1TNbSe$_2$ in the 
$\sqrt{13}\times\sqrt{13}~R30^{\circ}$ phase is ferromagnetic
 with a total spin magnetization of
$1\mu_B$ per cell and an absolute magnetization (sum of the modulus of
the magnetic moments on all atoms in the cell) of $2.2\mu_B$.

Before proceeding further, it is very instructive to consider the effect of the Hubbard
correction and of the stabilization of a  magnetic state on the
energetics of the different polytypes of single-layer NbSe$_2$.
Complete structural optimization within the GGA+U method leads to 
cell parameters in excellent agreement with experimental data even
 for what concerns the 2H non-magnetic high-symmetry structure
(see Tab. \ref{tab:Energy}), with essentially no change in its
electronic structure (see \cite{Supplemental})). 
The GGA+U corrects the GGA underbinding error.
The energy differences between
the non-magnetic 1T  and 2H high-symmetry structures is still large, but substantially
reduced, suggesting that the high-symmetry 1T phase cannot be easily
stabilized. The only effect of the GGA+U approximation on the
electronic structure of high-symmetry phase of single layer 1TNbSe$_2$ is to unmix Se p and
Nb d states at the Fermi level at zone center (see \cite{Supplemental}).
Most important, the total energy of the magnetic 
$\sqrt{13}\times\sqrt{13}~R30^{\circ}$ structure 
has a $1.6$ mRyd energy gain with respect to the 2HNbSe$_2$
high-symmetry structure. The 1TbSe$_2$ CDW phase is
comparable in energy with the high symmetry structure.
 Thermal or cinetic effects 
stabilize then one or the other at 500 K. 

In order to compare different approximations for the exchange
correlation functional on the electronic structure,
we stick to the the experimental cell parameter, as 
cell relaxation is anyway marginal. We  disentangle the
magnitude of the Hubbard term from the magnetic-induced structural
distortion by first calculating the electronic structure within the
GGA+U approximation on top of the 
GGA geometry. A ferromagnetic state is stabilized even in this
case with opening of an Hubbard gap between the majority
and minority bands having dominant d$_{z^2-r^2}$ character on the
central Nb atom, as shown in Fig. \ref{Fig:bands} ( see filled and open red dots
in the center panel). However, the electronic structure is still
metallic due to the residual hybridization between the d$_{z^2-r^2}$ central 
Nb and the nearest neighbours Se atoms. 

Structural optimization of internal coordinates within the GGA+U
approximation in the magnetic state leads to a substantial 
 distortion
of the $\sqrt{7}\times\sqrt{7}$ cluster around the central Nb
(i.e. the 6 Nb atoms in blue in Fig. \ref{Fig:struct_dist} , left).
If the GGA+U approximation with $U=2.95$ eV is performed at clamped nuclei,
i. e. it is carried out on the CDW structure obtained
at $U=0$ eV, the loss in total energy with respect to the fully optimized
structure at $U=2.95$ eV is $\approx 0.8$ mRyd/Nb atom, corresponding
to $\approx 30\%$ of the total energy gain by the CDW distortion
at $U=0$ eV (see Tab. \ref{tab:Energy}). 

The distortion leaves unshifted the central Nb atom, 
while the 6 nearest Nb atoms
move in-plane and apart from the central one and the out of plane Se atoms fills
the empty space reducing the distance with the Nb plane, as shown in
Fig. \ref{Fig:struct_dist} (center and right). 
The hybridization between the central Nb d$_{z^2-r^2}$ orbital and the 
nearest Se p-states is reduced and the band degeneracy is removed by
 opening an gap in the spectrum, as in a molecular Jahn-Teller effect. The
minority band with the largest Nb d$_{z^2-r^2}$ orbital component
becomes non-bonding (see Fig. \ref{Fig:bands}, right) .  The shift of the other
atoms not belonging to the $\sqrt{7}\times\sqrt{7}$ cluster, is minor,
so that overall the David-star CDW is preserved.
More quantitatively, using the experimental lattice parameter,
the distance between the central Nb atom in the star and
its neighbouring Nb is $\approx 3.157\AA$ at $U=0$ eV and 
$\approx 3.226\AA$ at $U=2.95$ eV. Similarly, the distance of between
Se atoms of the $\sqrt{7}\times\sqrt{7}$ cluster from the Nb-plane is
decreased from $\approx 1.838 \AA$ at $U=0$ eV to $1.788\AA$
at $U=2.95$ eV.
 
Thus, the insulating state in 1TNbSe$_2$ is not simply reached via a
standard Mott mechanism, but a cooperative lattice and magnetic
effect increases the Mott gap and amplifies the effect of the Hubbard interaction. 
Within GGA+U, single-layer 1TNbSe$_2$ is then a phonon-assisted spin-$1/2$ ferromagnetic
Mott insulator displaying a $\sqrt{13}\times\sqrt{13}~R30^{\circ}$
David-star CDW. 
\begin{table*}[b]
\begin{tabular}{l|c c c c |c c} 
\hline 
Structure & $\Delta E (U=0, {\rm Exp.})$ & $\Delta E (U=2.95,{\rm 
                                           Exp.})$  & $\Delta E (U=0,
                                                      {\rm Opt.})$ & $\Delta E (U=2.95,{\rm 
                                           Opt.})$ &$a(U=0)$& $a(U=2.95)$  \\ %
\hline 
2H high-symmetry             & $0.0$  &  0.0 & 0.0 & 0.0 & 3.473 & 3.451 \\ 
1T high-symmetry  & $+7.2$ &  +4.2 & +7.3 & +4.4 & 3.482 &3.464 \\ 
1T CDW    & $+3.0$ & -1.1  & +2.9 & -1.6 & 12.575 & 12.523 \\ 
\hline 
\end{tabular}
\caption{Energetics of single-layer NbSe$_2$ polytypes with respect to 
 the high-symmetry 2HNbSe$_2$ polytype, $\Delta E=E-E(2H)$, as a 
 function of $U$. The labels $\Delta E (U=0, {\rm Exp.})$, $\Delta E (U=2.95,{\rm Exp.})$
means that the internal coordinates have been optimized at fixed volume using 
the experimental lattice parameters and the GGA+U functional with $U=0$ and
$U=2.95$ eV respectively. The labels $\Delta E (U=0, {\rm Opt.})$ and
$\Delta E (U=2.95,{\rm Opt.})$ refers to complete structural optimization
of internal coordinates and volume using the GGA+U approximation with 
with $U=0$ and  $U=2.95$ eV respectively.  The energy units are mRyd per Nb atom and $U$
is expressed in eV. The values of the experimental (Exp.) and optimized (Opt.) cell 
 parameter are shown (in $\AA$). We use as experimental lattice
 parameter the in-plane one of the bulk, namely $a_{\rm Exp.}=3.44
 \AA$.}
\label{tab:Energy}
\end{table*}

It is worth to mention that our calculated gap is
approximately 0.18 eV,  smaller
than the one detected in STM ($0.3-0.4$ eV) and
 ARPES in Ref. \cite{Nakata}. The gap does not seems
to  increase anymore with $U$, as for $U> 3.1$ eV, the flat band
becomes higher in energy than the next two empty
bands\cite{Supplemental,Pasquier}.
After submisison of our work, it was
suggested \cite{Wehling} that the gap underestimation 
could be due to the lack of non-local exchange in the  
calculation. 
Beside this, other effects
could also affect the gap size like the presence
of a graphene substrate leading to a non-negligible charge transfer or
 band-bending  occurring in STM.
Nevertheless, it is important to underline that the valence
band electronic structure is in good agreement with experimental data despite the large
temperature broadening occurring in experiments\cite{Supplemental}. 

In this work, by using first principles calculations, we have shown that
single layer 1TNbSe$_2$ undergoes a CDW instability
with a $\sqrt{13}\times\sqrt{13}~R30^{\circ}$ reconstruction. 
However, the GGA leads to a 
substantially unstable 1TNbSe$_2$ structure with respect to the
2H polytypes. The GGA+U
approximation improves the stability of all 1T polytypes, explaining
the detection of 1TNbSe$_2$ in experiments. 
Moreover, it stabilizes a ferromagnetic state with a spin magnetization of
$1\mu_B$. Ferromagnetism occurs in the context of a flat d-band,
as in the ferromagnetic cuprate
LaBaCuO$_5$\cite{Eyert,Pickett}. However, while in the latter
the flat band is isolated and even the LSDA 
stabilizes a sizeable gapped ferromagnetic state, 
in 1TNbSe$_2$ the situation is more complicated by the non-negligible hybridization 
with Se states at zone center and the framework is more that
of a multiband Hubbard model. If the lattice is kept frozen at the
non-magnetic crystal structure, then
hybridization hinders the opening of a Mott gap 
and favours a metallic state (or a state with a negligible gap). 
Thus, even if to some extent the long-standing proposal of
Ref. \cite{Fazekas} based on a purely electronic mechanism 
partly applies to 1TNbSe$_2$, this is also
not the complete explanation as
the opening of the
gap occurs via a cooperative electronic and lattice (Jahn-Teller) effect, 
resulting
from  ferromagnetism and the subsequent Nb-Nb bond-softening
around the central Nb atom in the star.
Even if for small values of $U<3.1$ eV 
the size of the gap depends on the value of the Hubbard U
parameter (that in our work is calculated from first principles
self-consistently\cite{CococcioniSCF}), 
the distortion amplifies anyway the effect of the Hubbard
term by reducing the hybridization (the larger U, the larger the local
distortion). 

We believe that the cooperative enhancement of the 
Mott gap due to  electronic and
lattice degrees of freedom is revelant far beyond the family
of 1T dichalcogenides, but it is a genuine feature of 
correlated insulators in the presence
of strong Jahn-teller distortion. A similar mechanism
was suggested to be relevant in manganites and manganese oxides
\cite{Motome} or in the fullerides like K$_4$C$_{60}$ \cite{Fabrizio}. 
1TNbSe$_2$ is an ideal playground to study Mott Jahn-Teller insulators
in reduced dimension. 

We acknowledge IDRIS, CINES and TGCC and PRACE for high
performance computing resources, support from
the European Union Horizon 2020 research
and innovation programme under Grant agreement
No. 696656-GrapheneCore1 and from Agence Nationale de
la Recherche under the reference No. ANR-13-IS10-0003-
01. We acknowledge M. Casula for useful discussions.


\end{document}